\documentstyle[12pt]{article}
\begin {document}
\setcounter{page}{0}

\title{Spin effects in heavy quarkonia}
\author{L. Motyka and K. Zalewski\thanks{Also at the Institute of Nuclear
 Physics, Krak\'ow, Poland}\\
 Institute of Physics, Jagellonian University, Krak\'ow, Poland}
 \maketitle

\begin{abstract}
{A model based on a nonrelativistic potential supplemented by spin dependent
terms of the Breit-Fermi type is described. This model reproduces the known
masses of the $\overline{b}b$ quarkonia within $1$MeV and the known masses of
the $\overline{c}c$ quarkonia within $5$MeV. Many predictions are made, some of
them, e.g. for the fine splittings of the high $L$ states of the
$\overline{b}b$ quarkonia and of the mass spectrum of the $\overline{b}c$
quarkonia, significantly different from previous predictions made by other
authors.}
\end{abstract}

\section {Introduction}

For the hydrogen atom the nonrelativistic Schr\"odinger equation yields a good
first approximation for the energy levels. The spin effects are responsible
only for the fine and hyperfine structure of the levels. In heavy quarkonia, as
well, the spin effects yield the fine and hyperfine splittings, but there are
two important complications. In the hydrogen atom fine splittings are
negligible compared to the energy differences between the nonrelativistic
levels and hyperfine splittings are negligible compared to the fine splittings.
In heavy quarkonia the fine and hyperfine splittings are of the same order.
They are smaller than the energy intervals between the unsplit levels, but
certainly not negligible. Therefore, the comparison of nonrelativistic
calculations with experiment is ambiguous. In the hydrogen atom the velocity of
the electron in the ground state is about $0.01 c$, thus the nonrelativistic
approximation is a well justified starting point. In heavy quarkonia the quarks
are much faster. The root mean square velocity is, in the ground state of the
$\overline{b}b$ system, about $0.3 c$ and in the ground state of the
$\overline{c}c$ system about $0.5 c$. The $c$-quark in the $\overline{b}c$
system is probably even faster than that in the $\overline{c}c$ system. It may
seem, therefore, that the nonrelativistic approximation is suitable only for
crude estimates. Nevertheless, many authors have found that nonrelativistic
models give amazingly successful predictions for the spin averaged energy
levels of heavy quarkonia. Including perturbatively spin effects one can also
reproduce very satisfactorily the fine and hyperfine structure. Many references
to work of this kind can be found in the review paper \cite{BES}. A
simple-minded interpretation is that most of the relativistic effects is
somehow taken into account by fixing the parameters of a nonrelativistic model
to fit the experimental data -- they just renormalize the parameters of the
nonrelativistic potential. It could be argued that instead of the poorly
understood potential models it is better to use QCD sum rules (cf. e.g.
\cite{NAR}) or lattice methods (cf e.g. \cite{DAV}). At present, however, these
methods have rather large uncertainties. In the present paper we describe a
model \cite{MZ1}, \cite{MZ2}, which gives a remarkably good description of the
mass spectra below the strong decay thresholds, i.e. below $10.558$ GeV for the
$\overline{b}b$ quarkonia and below $3.729$GeV for the $\overline{c}c$
quarkonia. This makes it plausible that also its predictions for the
$\overline{b}c$ quarkonia below $7.144$ GeV ($7.189$ GeV) will not be far off
the mark. The mass limit in brackets is applicable to quarkonia, which, because
of angular momentum and parity conservation, cannot decay into two pseudoscalar
mesons. In these ranges there are $34$ mass states for the $\overline{b}b$
quarkonia (out of that nine are known), $16$ mass states for the
$\overline{b}c$ quarkonia and 8 states for the $\overline{c}c$ quarkonia (out
of that six are known). We consider as known only the mass states included as
firmly established in the 1986 Particle Data Group Tables \cite{PDG}.

\section {Nonrelativistic potential}
Our model contains six free parameters -- the quark masses

\begin{equation}
m_b = 4.8030\mbox{GeV},\;\;\;\;\; m_c = 1.3959\mbox{GeV},
\end{equation}
the coupling constant $\alpha_s$ at the $m_c$ mass scale

\begin{equation}
\alpha_s(m_c) = 0.3376
\end{equation}
and the three parameters of the nonrelativistic potential

\begin{equation}
a = 0.32525,\;\;\;b=0.70638\mbox{GeV}^\frac{3}{2},\;\;\; c = 0.78891\mbox{GeV}.
\end{equation}
Of course we do not claim to have fixed the quark masses with an accuracy of
four digits etc. All these digits are necessary, however, in order to reproduce
our results. The potential is spherically symmetric and reads

\begin{equation}
V(r) = m_Q + m_{\overline{Q}} - c + b\sqrt{r} -  \frac{a}{r}.
\end{equation}
The eigenvalues of the corresponding Schr\"odinger equation are interpreted as
the centres of mass of the spin one multiplets. For the $\overline{b}c$
quarkonia this corresponds to the case, when the singlet-triplet mixing is
switched off.

Comparison with experiment for the $\overline{b}b$ quarkonia can be made in
five cases. The agreement is within $1$ MeV. A rough estimate of the
relativistic correction gives about $60$ MeV. Thus, almost all this correction
must reduce to a redefinition of the parameters of the potential. For the
$\overline{c}c$ quarkonia comparison with experiment can be made in three
cases. The dicrepancies are within $4$ MeV. Since the relativistic corrections
here should be much larger than in the previous case, the good agreement is
even more surprising than before. It is instructive to compare our results with
those of Gupta and Johnson \cite{GUJ}. They are using a very different model,
but their fit is about as good as ours. In fact it is somewhat better for the
$\overline{c}c$ system, though somewhat worse for the $\overline{b}b$ system.
In spite of the agreement within a few MeV between the two models for the
$\overline{b}b$ and $\overline{c}c$ systems, the predictions for the
$\overline{b}c$ quarkonia differ significantly. We expect higher masses, by
about $40$ MeV for the $S$ states and by $16$ MeV for the $P$ multiplet. Thus,
a good fit for the $\overline{b}b$ and $\overline{c}c$ is compatible with a
poor fit for the $\overline{b}c$ quarkonia. Of course, we do not know yet,
whose fit is poor, but both fits cannot be good.

\section{Hyperfine splittings}
The hyperfine splittings are usually described using the Breit-Fermi
interaction, which in the present case takes the form (cf. e.g. \cite{BUH})

\begin{equation}
H_{HF} = \frac{32 \pi \alpha_s}{9m_Qm_{\overline{Q}}} \left(\vec{S}_Q \cdot
\vec{S}_{\overline{Q}} - \frac{1}{4} \right) \delta(\vec{r}).
\end{equation}
In the first approximation of the Rayleigh-Schr\"odinger perturbation theory
the corresponding energy splitting is

\begin{equation}
\Delta E_{HFS} = \frac{32 \pi \alpha_s}{9m_Qm_{\overline{Q}}}|\psi(0)|^2.
\end{equation}
In this approximation, the splitting occurs only for the $S$ states, because
for the states with $L \neq 0$ the wave function $\psi$ vanishes, when the
distance between the two quarks goes to zero. The measured splitting of the
$1S$ state of the $\overline{c}c$ system yields the value of $\alpha_s$ (2).
Once this is chosen, it is possible to predict any other hyperfine splitting,
but there is no data to compare the predictions with. Some indirect support is
obtained by calculating the coupling constant $\alpha_s$ at the mass of the
$Z^0$ --- the result $\alpha_s(M_{Z^0})= 0.115$ is very acceptable --- and from
the calculation of the leptonic decay widths of heavy quarkonia \cite{MZ2}. It
is instructive to compare our prediction for the hyperfine splitting of the
$1S$ state in the $\overline{b}b$ system ($56.7$ MeV) with the predictions of
the sum rules \cite{NAR} --- $63 \pm 30$ MeV --- and of the lattice approach
\cite{DAV} --- $60$ MeV with a large, but unspecified, error. Our agreement
with the mean values is good, but in view of the large uncertainties of the
more rigorous approaches, it is premature to draw any conclusions.

\section{Fine splittings and mixing}
The interaction terms responsible for the fine splittings are (cf. e.g.
\cite{EQU}, \cite{GER})

\begin{eqnarray}
V_A(r) &=&   \frac{m_Q^{-2} - m_{\overline{Q}}^{-2}}{4}\left( -\frac{1}{r}
\frac{\partial V }{\partial r} + \frac{8\alpha_s}{3r^3}\right) \vec{L} \cdot
\left(\vec{S}_Q - \vec{S}_{\overline{Q}}\right),\\
V_{LS}(r)   &=& \left[\frac{m_Q^{-2} + m_{\overline{Q}}^{-2}}{4}\left(
-\frac{1}{r} \frac{\partial V}{\partial r} + \frac{8\alpha_s}{3r^3}\right) +
\frac{4 \alpha_s}{3 m_Q m_{\overline{Q}} r^3}\right]\vec{L} \cdot
\left( \vec{S}_Q + \vec{S}_{\overline{Q}}\right),\\
V_T(\vec{r}) &=& \frac{4\alpha_s}{3m_Qm_{\overline{Q}}r^5} \left[3(\vec{S}_Q
\cdot
\vec{r})
(\vec{S}_{\overline{Q}} \cdot \vec{r}) -r^2 \vec{S}_Q \cdot
\vec{S}_{\overline{Q}}\right].
\end{eqnarray}
In these formulae the only undefined quantity is the coupling constant
$\alpha_s$, which should be running, i.e. depending on $r$. We propose the
following procedure to define this function. The general formula is modelled on
the well-known one-loop expression for $\alpha_s(p)$ with the replacement of
the momentum scale $p$ by $\frac{1}{r}$. We introduce tildes above $\alpha_s$
and $\Lambda$, when considered as functions of $r$,  in order to avoid
confusion with the corresponding quantities considered as functions of
momentum. Thus

\begin{equation}
\tilde{\alpha}_s(r) = \frac{12\pi}{33 - 2n_f}
\frac{\left(\tilde{\Lambda}^{(n_f)}\right)^2r^2-1}{\log\left(
\left(\tilde{\Lambda}^{(n_f)}\right)^2r^2\right)}.
\end{equation}
Here the number of flavours $n_f$ equals three for $r>m_c^{-1}$, equals four
for $m_b^{-1} < r < m_c^{-1}$ and equals five for $r<m_b^{-1}$. The region $r>
m_s^{-1}$ is of little interest for the present problem, therefore, we keep
$n_f=3$ also for large values of $r$. The numerator is introduced in order to
compensate the zero of the denominator at $r = \frac{1}
{\tilde{\Lambda}^{(n_f)}}$. Its exact shape is of little importance. The
parameter $\tilde{\Lambda}^{(4)}$ is obtained from the two conditions

\begin{eqnarray}
\tilde{\alpha}_s\left(\frac{1}{m_c}\right) = \alpha_s(m_c),\\
\tilde{\alpha}_s\left(\frac{1}{m_b}\right) = \alpha_s(m_b).
\end{eqnarray}
These two equations give slightly different values of the parameter
$\tilde{\Lambda}^{(4)}$ and we take the geometrical mean of the two results.
However, the two solutions are so close to each other, that taking the
arithmetic mean instead of the geometric one would make no difference at our
level of precision. Then, $\tilde{\Lambda}^{(3)}$ and $\tilde{\Lambda}^{(5)}$
are calculated from the conditions that the function $\tilde{\alpha}_s(r)$
should be continuous at $r=m_c^{-1}$ and at $r = m_b^{-1}$. Interpreting
$\alpha_s$ in the formulae for the potentials $V_A,\;V_{LS}$ and $V_T$ as the
function $\tilde{\alpha}_s(r)$, we obtain the explicit form of the operator
responsible for the fine splittings of the mass levels of the heavy quarkonia.

When calculating the splittings of the $\overline{b}b$ quarkonia, we find very
good agreement with experiment for the splittings of the $1P$ and $2P$ states.
The accuracies of our results are all within $1$MeV. Most other models,
however, are also successful in these predictions, typical errors being within
$5$ MeV. Also for the $3P$ states, where there is no data to compare with, our
predictions are close to those of other authors. A qualitative difference
between the predictions appears, however,  for the higher $L$ states. For
instance, Kwong
and Rosner \cite{KRO} predict a splitting of about $1$ MeV for the $F$-states
and almost no splitting for the $G$-states, while we find a splitting of about
$13$ MeV for the $1F$ state and of about $10$ MeV for the $1G$ state. Thus, the
high $L$ states will be of great interest for the comparison of models. For the
$\overline{c}c$ systems our predictions for the splitting of the $1P$ state
agree with experiment within about $5$ MeV. For other states we are in rough
agreement with other models, except for the ${}^3D_1$ state, where we predict
a down shift of about $20$ MeV, while other models find much smaller shifts.

For the $\overline{b}c$ system one can calculate the singlet-triplet mixing due
to the operator $V_A$. These angles are very sensitive to the details of the
models and comparing the results of our paper and of the papers \cite{EQU} and
\cite{GER} one sees that no two sets of predictions agree. According to our
calculation the sines of the mixing angles for the $1P,\;2P$ and $1D$ states
are respectively: $0.374,\;0.385$ and $0.244$. Eichten and Quigg \cite{EQU}
find almost no mixing for the $1P$ state, while Gershtein and collaborators
\cite{GER} find the biggest mixing for the $1D$ state.

Another interesting mixing effect is due to the operator $V_T$. This involves
pairs of states differing by two units in orbital angular momentum. The mixing
angles are very small --- less than $10^{-3}$ --- and the effect on the mass
values is negligible. This effect, however, enhances considerably the leptonic
decays of the quarkonia with angular momenta $L \geq 2$.

 \end{document}